# Evolution of clusters in energetic heavy ion bombarded amorphous graphite-II: formation and fragmentation phenomena

A. Qayyum[1], B. Ahmad[1], M.N. Akhtar[1], and **Shoaib Ahmad**[2,*]
[1]PINSTECH, P.O. Nilore, Islamabad, Pakistan
[2]National Centre for Physics, Quaid-i-Azam University Campus, Shahdara Valley, Islamabad, 44000, Pakistan
*Corresponding author: Email: sahmad.ncp@gmail.com

## Abstract

A study has been conducted into the mechanisms of evolution of clusters and their subsequent fragmentation under energetic heavy ion bombardment of amorphous graphite. The evolving clusters and their subsequent fragmentation under continuing ion bombardment are revealed by detecting various clusters in the energy spectra of the Direct Recoils (DRs) emitted as a result of collisions between ions and surface constituents. The successive DR spectra reveal that the energetics of C−C bond formation as well as any subsequent fragmentation can be related to the processes of energy dissipation in a cylindrical volume of a few Å surrounding the ion path. The dependence of $C_m$ formation or $C_m \rightarrow C_{m-2} + C_2$ fragmentation is seen to be a function of the ionic stopping powers in this cylindrical volume.

## 1. Introduction

A whole range of carbon cluster techniques have been employed following the demonstration of the production of carbon clusters $C_m^+$ with *m* ranging from 1 to about 200 [1] followed by the discovery of Buckminsterfullerene in the laser ablated graphite plumes in controlled environment [2]. Just as the means of energy deposition vary from laser-quanta [1,2], arc discharge [3] to high energy density electron beams [4,5] and ions from keV to GeV [6-11], similarly a wide variety of solids from graphite to polymers has been experimented with. In addition to the attempts of achieving gram quantities of $C_{60}$ and $C_{70}$ with these techniques, the possible formations of fullerenes and higher clusters and nanotubes have also been actively investigated. Energetic ion irradiation has been investigated as a tool of cluster production in polymers [6-8]. The high pressure Chromatography of pyrolytic graphite samples [9] and sugar molecules [10] irradiated with MeV and GeV ions have shown that $C_{60}$ is being produced in these diverse carbon containing materials.

We had earlier reported the observations of clusters ranging from the linear chains and rings to fullerenes in the direct recoil energy spectra from amorphous graphite bombarded with 100 keV Kr[+] and Xe$^+$ ions [11]. Amorphous graphite was chosen to ensure a crystal structure-less medium where carbon atoms could undergo ion induced sequences of bond breaking and re-bonding to produce complex structures.

The constituents of a surface recoiling in a binary collision with an incident ion are the primary knock-ons of radiation damage theory also known as the Direct Recoils -DRs. A DR carries a characteristic energy which is a function of the target to projectile mass ratio ($m_2/m_1$), angle of recoil $\theta_{DR}$ and the bombarding energy $E_0$. The energy of graphite atoms of mass $m_2$ recoiling at angle $\theta_{DR}$ is given by [12]

$$E_{DR} = 4\frac{m_1 m_2}{(m_1 + m_2)} 2E_0 \cos^2 \theta_{DR}. \tag{1}$$

Where $m_1$ and $E_0$ are the mass and energy of the projectile. The differential recoil cross section $d\sigma_r/d\Omega$ in the Lab frame can be worked out from differential scattering cross section $d\sigma(\zeta)/d\omega$ in the centre of mass (C.M.) system for C.M. angle _ [12]

$$d\sigma_r/d\Omega = 4\sin(\zeta/2)\, d\sigma(\zeta)/d\omega \tag{2}$$

by using Kr−C potential and following [13], we have

$$d\sigma(\zeta)/d\omega = \frac{m_2 E_0}{(m_1 + m_2)} \frac{3.05 Z_1 Z_2}{\left(Z_1^{1/2} + Z_2^{1/2}\right)^{2/3}} \frac{\pi^2 (\pi - \zeta)}{\zeta^2 (2\pi - \zeta)^2 \sin \zeta} \tag{3}$$

While calculating $d\sigma_r/d\Omega$ for selective values of $\theta_{DR} = (\pi/2 - \zeta/2)$ a minimum direct recoil energy $E_{DR}(min) \sim 10$ eV corresponding to $\theta_{DR} \approx 89°$ is to be established. This minimum is necessary to dislodge a carbon atom from the graphite surface.

These Direct Recoils being the primary events of the energetic projectile-target interaction subsequently initiate collision cascades in which the energy is shared with other neighbours in the solid. Most of the sputtering yield is due to ejections from solids upon interaction of these collision cascades with the surface. Sputtering yield theories [14,15] predict that the yield S or the total flux of atoms sputtered in all directions and energies for unit flux of incident ions is directly proportional to the deposited energy $F_D$ and inversely to the surface binding energy $E_b$ of the target atoms as $S \propto F_D/E_b$. The density of energy deposition at the surface $F_D$ can be further estimated [14] by using the nuclear stopping cross section $S_n(E_0)$; $F_D \approx v\eta S_n(E_0)$ where $v = v(m_2/m_1)$ is a target to projectile mass dependent parameter and $\eta$ is the target surface density. Using the TRIM96 code [16] nuclear stopping cross section $S_n(E_0)$ is estimated between 40 and 160 [eV/Å] for 100 keV Ar$^+$; Kr$^+$-and Xe$^+$ ion beams. Surface binding energy of carbon atoms $E_b$ can be estimated ~ 2−3 times $E_{C-C}$; where $E_{C-C}$ is a single C−C bond energy ≈ 3.6 eV.

The DRs have well defined energies, trajectories and points of origin whereas, the sputtered particles have broad energy distribution peaked at $E_b/2$ [15] and result from various trajectories and origins. The time scales of the two interactions are also widely different; individual DR events take $10^{-14}$ to $10^{-15}$ s, while the cooling down of the heavy ion-induced collision cascade typically takes up to $10^{-11}$ s. The two types of projectile-initiated collision events are distinct and therefore, the energy spectrum of DRs can positively discriminate against the low energy sputtered particles. In our experimental set up we detect the direct recoils while the sputtered particles are discriminated against in the detection process. However, both of the mechanisms are intricately involved in the ion-induced clustering mechanisms in the solid and will be discussed later in Section 4.

## 2 The experiment

### 2.1 The experimental setup

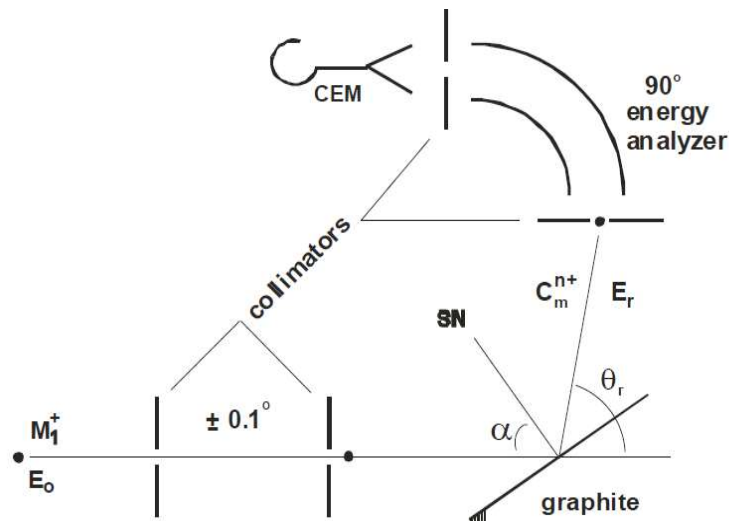

An indigenously designed and fabricated PINSTECH ion accelerator, a 250 keV heavy ion facility has been used for the experiments. $Ar^+$; $Kr^+$ and $Xe^+$ beams of > 1 mm diameter and energy between 50 to 250 keV can be delivered to a target 2 meters from the end of the Accelerator tube. The facility is equipped with a hollow cathode duoplasmatron operating at pressure $\sim 10^{-2}$–$10^{-3}$ mbars, with the accelerator delivering a few $\mu$A collimated beam with $\sim \pm 0.1°$ divergence on the target. The pressure in the target chamber $\sim 10^{-7}$–$10^{-8}$ mbars.

The experimental setup is shown in Figure 1 where the beam and the recoil particles' collimators with less than $\pm 0.1°$ divergence are shown along with retractable 90° electrostatic energy or a momentum analyzer. Experiments are performed with target chamber at pressures $10^{-7}$ mbars maintained with an ion pump. A Channel Electron Multiplier (CEM) is used for cluster detection. CEM with a typical gain of 5 x $10^7$ feeds the charged recoil data to a PC via a rate meter and a Hydra Data Acquisition unit. The energy analyzer's condenser plate potential is increased in variable steps through a function generator (Philips PM 5138). The resolution of the EEA is $\sim$ 0.01 with 0.8 mm entrance aperture for EEA. Solid angle $d\Omega = 6 \times 10^{-6}$ [st. rad.]. The electrostatic analyzers can allow detection of up to tens of keV heavy recoiling clusters. Although momentum analysis of clusters is desirable to unambiguously characterize the $m/q$ values but the required magnetic fields become unrealistically large for experimental arrangements like ours. For example, in case of $\theta_{DR}= 79.5°$, a large magnet is needed with $B_0\rho$ = 4 [T-m] for resolving clusters e.g., $C_{60}$. For $\theta_{DR}$= 87.8°, momentum analysis has been performed with a magnet with $B_0\rho$ = 0.06 [T-m] and the results compared with those from electrostatic analysis in Figure 2. This analyzer is appropriate only for smaller recoil angles.

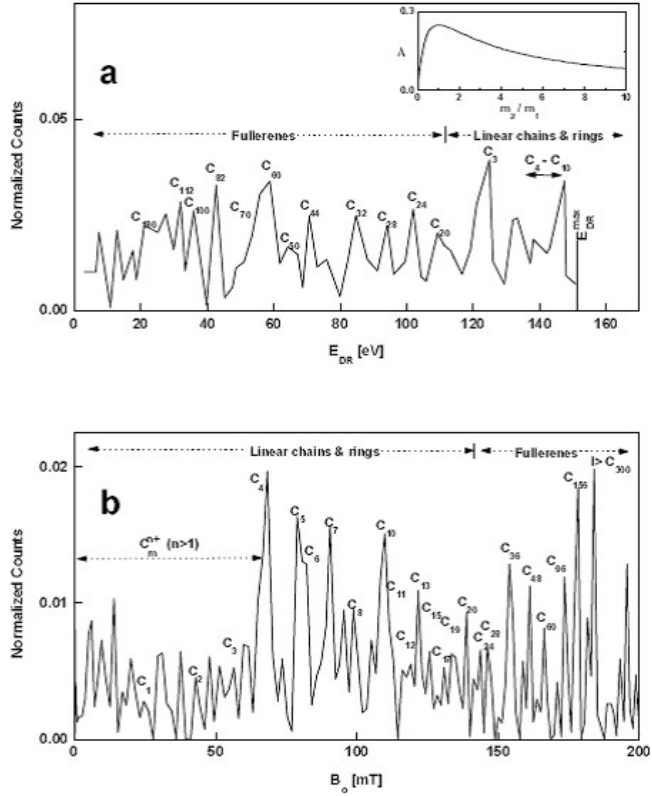

Fig. 2. A comparison between the energy and momentum analyses is shown in Figures 2a and 2b, respectively with 100 keV $Kr^+$. Both of the spectra are obtained from almost similar physical conditions. The mass analysis favours smaller masses while the heavier fullerenes are grouped at the end of the spectrum in Figure 2b. The situation is exactly opposite in the case of recoil energy spectrum shown in Figure 2a. In the $E_{DR}$ spectrum the clusters with masses $m_2 > m_1$ are well spread out while those with $m_2 < m_1$ are squeezed; see the inset of Figure 2a.

## 2.2 Electrostatic versus momentum analyses for the DRS of ion-induced clusters

A summary of the comparative features of Direct Recoil Spectroscopy by using electrostatic energy selection vs. the momentum analysis is shown in Table 1, electrostatically analyzed DR spectrum is contained within the range of recoil energies from 0 to $E_{DR}^{max}$ and the whole spectrum is always produced. There is a well defined $E_{DR}^{max}$ which can be imparted to the target constituents by the incident ions of energy $E_0$ at the angle $\theta_{DR}$. This happens at $m_2 = m_1$. The $E_{DR}$ spectrum is spread out in such a way that the clusters with masses $m_2 > m_1$ are well spread out while those with $m_2 < m_1$ are squeezed; see the inset of Figure 2a where the ratio $\Lambda = (m_1 m_2)/(m_1 + m_2)^2$ of equation (1) is plotted against $m_2/m_1$.

Due to the nature of recoil kinematics in equation (1), certain lower mass peaks can overlap with heavier ones for a given projectile. Therefore, different projectiles are required for unambiguous identification of respective peaks. The momentum analysis on the other hand, provides a continuous spectrum of gradually increasing masses as a function of the analyzing field. Multiply charged recoils can appear at half the respective fields and in some cases, overlap with lighter particle's singly charged peaks The deflected projectiles within the target can also initiate recoils but with higher or lower energies than those expected for a particular ion-cluster combination. This effect can be seen in the broadening of the $E_{DR}$ peaks for respective clusters. This effect is reproduced by both of the analyzing techniques.

Figure 2 provides a comparison between the energy and momentum analyses shown in Figures 2a and 2b, respectively. The data is obtained by using 100 keV Kr$^+$ beam. Both of the spectra are obtained from similar physical conditions of the target, ion irradiation, beam intensity and the recoil angle. The mass analysis favours smaller masses and the heavier fullerenes are grouped at the end of the spectrum in Figure 2b. The situation is exactly opposite in the case of recoil energy spectrum shown in Figure 2a.

It can be seen from the comparative spectra of Figures 2a and 2b that the momentum analyzer provides a well resolved spectrum for smaller clusters ($C_m < C_{36}$) but a tightly squeezed spectrum for $C_{36}$ to $C_{70}$. On the other hand the energy spectrum is well spaced out for $C_{20}$ to $C_{240}$. Our requirements are that; the entire cluster spectrum be available. Therefore, in our experimental conditions and cluster identification requirements, energy analysis is preferred over the corresponding momentum analysis.

| Electrostatic Analysis | Momentum Analysis |
|---|---|
| 1. The entire $DR$ spectrum is contained within the range of direct recoil energies from $0 \to E_{DR}^{max}$. | 1. No clear cut upper limit on $m/q$ values. The maximum magnetic field of a specific target will determine $B_0^{max}$ and therefore, the highest $m/q$. |
| 2. Some overlap of peaks may occurs for certain ion-target combinations e.g. $C_1$ and $C_{50}$ for Kr$^+$−C and $C_2$ and $C_{60}$ for Xe$^+$−C. | 2. No peak overlap; clusters $C_m$ appear in increasing order of $m$ as a function of the analysing field $B_0$. |
| 3. Multiply charged $C_m^{n+}$ ($n > 1$) of lower masses ($m \leq 20$) may overlap with the fullerene's singly charged peaks. | 3. Multiply charged fullerenes $C_m^{n+}$ ($m > 20; n > 1$) peaks may overlap with those from the singly charged lower mass cluster. |
| 4. The main fullerenes $C_m$ with ($36 \leq m \leq 106$) are well spread out while the lighter ($m < 36$) and the heavier ones ($m > 106$) are squeezed within $\sim 20-30\%$ of the entire range. The energy resolution is however, same for a $C_m$ for all C all $C_m$ irrespective of their mass. | 4. The smaller clusters $C_m$ ($m \leq 20$) are well spread out with increased resolution while the entire fullerene range is squeezed at the higher $B_0$ end with gradually reducing mass resolution. |

Table 1. A summary of the characteristic features of Direct Recoil Spectroscopy (DRS) with the electrostatic and momentum analyses.

## 3 Results

The results are presented for the detector count rate $I_D$ normalized by the ion current $I_{ion}$, irradiated area $\delta x$ and the solid angle $d\Omega$ as $dP/d\Omega = I_D/((I_{ion}/\delta x)d\Omega)$ [counts/ion/cm$^2$/st. rad.]. This normalized detector count rate $dP/\Omega$ is plotted as $P_r$ in the energy spectra as a function of energy of direct recoils $E_{DR}$. $P_r$ is the cross sectional area per unit solid angle $d\Omega$ of one incoming projectile to produce a particular target particle (monomer or a multimer) ejection in a direct recoil at a given recoil angle. The experimental uncertainty is ~ 20% and is dependent upon the ion current, incidence angle and the dose measurements. The differential recoil cross section $d\sigma_r/d\Omega$ is obtained from the data by using $d\sigma_r/d\Omega = (1/\eta)\, dP/d\Omega$; where $\eta = \varrho\delta x$; $\varrho$ is the target surface density. The peak intensity from the $dP/d\Omega$ ($\equiv P_r$) vs. $E_{DR}$ for a particular cluster $C_m$ can then converted into the differential recoil cross section $d\sigma_r/d\Omega$. The experimental setup is such that the ion incidence angle at the surface $\alpha = 80°$ so that the irradiated region is a solid wedge with wedge angle of 10° between the ion path and the target surface. The adjacent sides are approximately equal to the ion range i.e., between 700−1300 Å. The side opposite to the wedge angle is the maximum depth from the surface of the irradiated region and is 120−230 Å. Therefore, one can expect Direct Recoils initiated from the surface as well as sub-surface recoils to be ejected. Those recoils that are due to the deflected trajectories will also be present in the energy spectra.

### 3.1 Results from $\theta_{DR} = 79.5°$

The first batch of experiments were performed at $\theta_{DR}$ =79.5° where E$_{DR}$ in case of Ar$^+$−C for $C_1^+$ is 2.36 keV and 0.66 keV for $C_{60}^+$ at 100 keV bombardment. The respective values are 1.45 and 1.42 keV for Kr$^+$−C. We can see from the results in Figure 3 that in addition to m> 1 peaks (i.e., those belonging to carbon clusters $C_m^+$) the contributions due to multiply charged monomers $C_1^{n+}$ are also significant where n > 1. In case of Ar$^+$−C (Fig. 3a), $C_{50}^+$ and $C_{70}^+$ (if these are present) will coincide with n = 3 and 4 peaks, respectively. Change of projectile therefore, is essential for resolving their respective contributions. The $C_{60}^+$ peak, however, is clearly distinguishable. Other clusters have also been observed and peaks for m = 7 to 120 can be seen. Figure 3b is for $Kr^+$−C at the same recoil angle and energy as of $Ar^+$−C. The $C_{60}^+$ peak can be identified clearly and is 6 times more probable for $Kr^+$−C than in the case of $Ar^+$−C. It can be seen that $C_{70}^+$ is not significantly present while definitive conclusion about its presence needs comparison with the data from $Xe^+$−$C_1$. The essential feature of variation of projectile mass is to evaluate the relative contributions of $C_m^{n+}$ with m and n > 1 and to distinguish between clusters and multiply charged monomers.

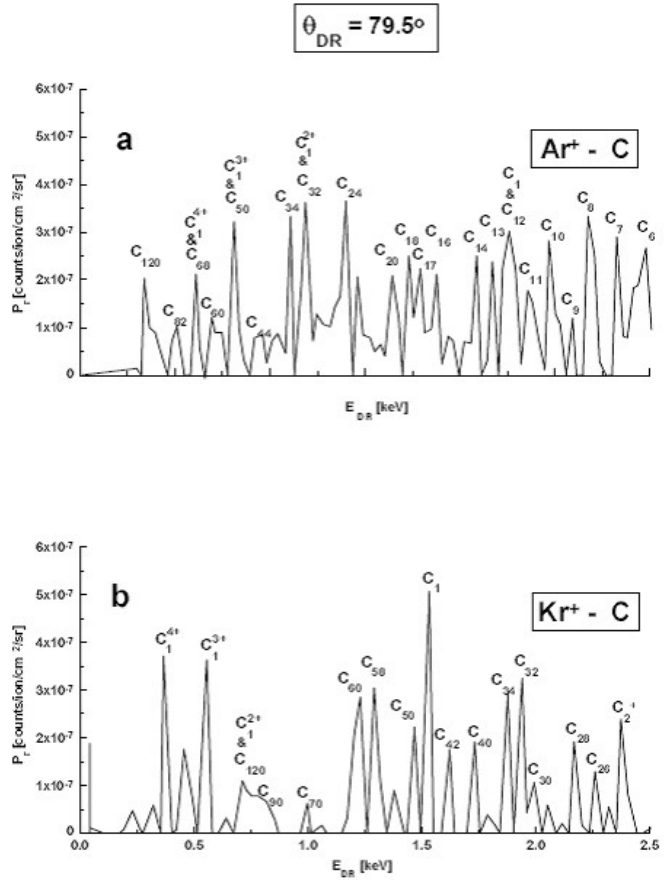

Fig. 3. a) Spectra from 100 keV Ar$^+$−C at $\theta_{DR} = 79.5°$. b) is for Kr$^+$−C at the same recoil angle and energy as of Ar$^+$−C. In addition to $m > 1$ peaks belonging to carbon clusters C$_m^+$, the contributions due to multiply charged monomers C$_1^{n+}$ are also significant where $n > 1$.

## 3.2 Results from $\theta_{DR} = 87.8°$

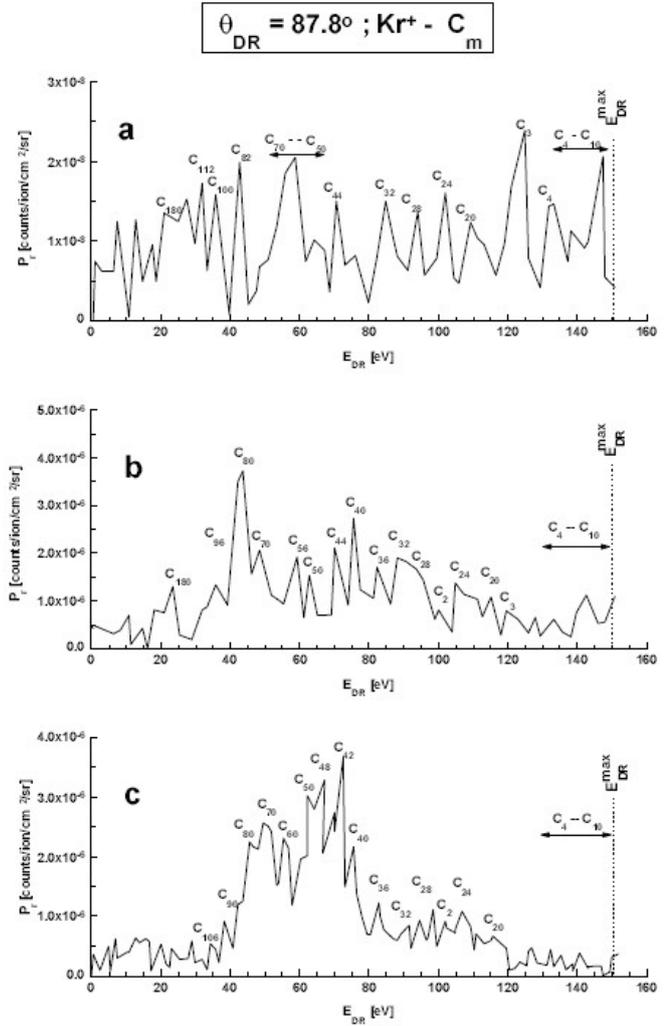

To supplement as well as clarify various experimental observations of direct recoil spectra at $\theta_{DR} = 79.5°$, we increased the recoil angle significantly to $\theta_{DR} = 87.8°$. It may be pointed out that the maximum recoil angle $\theta_{DR}^{max} = 90°$. At this angle $E_{DR} \to 0$. At $\theta_{DR} = 90°$ the spectra for the direct recoils of negligible energies ($E_{DR} \to 0$) merge with that due to the sputtered flux ejected by the collision cascades generated by the energetic projectiles. These sputtered particles have energies which are of the order of surface binding energies i.e., few eV. This implies that much smaller energies can now be imparted to carbon monomers and clusters at $\theta_{DR} = 90°$ with enhanced cross sections.

At $\theta_{DR} = 87.8°$ for $Kr^+$−C at 100 keV bombardment energy, the range of direct recoil energies to various $C_m^+$ is between 48 and 147 eV which is about 22 times less than their corresponding values at $\theta_{DR} = 79.5°$. In case of $Xe^+$−C, these recoil energies are even lower. This large recoil angle arrangement has two important aspects for the detection of charged clusters;

**Fig. 4.** Shows the Direct Recoil energy spectra at $\theta_{DR} = 87.8°$ from $Kr^+$−C at 100 keV. Carbon clusters $C_m^+$ are seen recoiling with characteristic energies $E_{DR}$. Figures 4a, b and c present the dynamic behavior of the formation and fragmentation of various clusters. $E_{DR}^{max}$ is the maximum transferable energy to a cluster in a direct recoil calculated from equation (1) and presented in the inset of Figure 2a.

a) due to much smaller recoil energies, the differential cross section is enhanced for all 3 projectiles, by at least an order of magnitude, and

b) the monomer multiplicity is considerably reduced as slower multiply charged ions pick up electrons while leaving the surface.

Thus we have cluster enhancement and monomer multiplicity reduction mechanisms operating simultaneously at $\theta_{DR}$=87.8° and this compares favourably with the cluster recoil emissions $\theta_{DR}$ =79.5°. Figure 4 shows the energy spectra of direct recoils from 100 keV Kr$^+$ ion bombarded graphite at recoil angle $\theta_{DR}$ =87.8°. Three consecutive spectra are shown with 4 Å beam incident at grazing incidence $\alpha \approx 80°$. The first spectrum (Fig. 4a) is taken after a dose of 2.5x10$^{14}$ ions. In addition to a broad peak around $C_{60}^+$ the entire fullerene range is present with higher m as well as the lighter ones including those with m < 36 and the linear/chain structural combinations (m = 1 to ∼ 10). The next spectrum (Fig. 4b) has $C_{80}^+$ as the dominant species while other cluster especially the fragments $\boldsymbol{C_{m-2}^+}$ resulting from $C_m \rightarrow C_{m-2} + C_2$ e.g., $C_{80}^+$, $C_{56}^+$, $C_{44}^+$, $C_{40}^+$ are conspicuous by their relative abundance. The lower order fullerenes have increased their share of the total yield. $C_{70}^+$ and $C_{50}^+$ are present but $C_{60}^+$ is not significantly present. The gradual building up of the $C_{50}^+$ and its fragments ($C_{48}^+$, $C_{46}^+$, $C_{42}^+$, $C_{40}^+$) can be seen from Figure 4c. A well defined peak $C_{60}^+$ for which compares well with those due to $C_{70}^+$, $C_{80}^+$, $C_{82}^+$ and $C_{50}^+$. The smaller clusters are present but their total yield is much smaller than that of the higher (i.e., > $C_{50}^+$) fullerenes.

Our earlier results from Figure 3b have indicated that in case of Kr$^+$−C, $C_{70}^+$ is least probable and $C_{50}^+$ if present in the spectrum, overlaps with m = 1 peak. To further clarify the presence or otherwise, of $C_{50}^+$, deductions from relative contributions of higher charged states of $C_1^{n+}$ (n >1) have shown that $C_{50}^+$ is also not significantly present. Thus the conclusion can be drawn from the interpretation of the persistent broad peak present in all the 3 spectra of Figures 4a, 4b and 4c is that it is due to $C_{60}^+$. The relative recoil probabilities seem to depend on the ambient irradiation conditions prevalent at the point of emission.

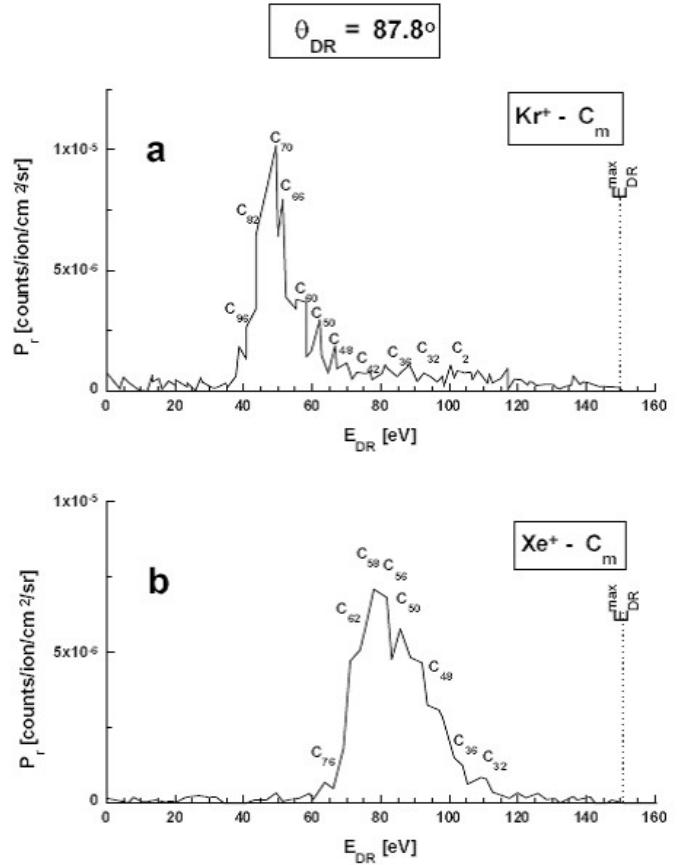

Fig. 5. Figure 5a is for 100 keV Kr$^+$−C and is dominated by C$_{70}^+$ and its fragments C$_{68}^+$ and C$_{66}^+$ along with shoulders of C$_{96}^+$, C$_{82}^+$, C$_{60}^+$, and C$_{50}^+$. The Xe$^+$−C spectrum of Figure 5b is likewise dominated by C$_{60}^+$ and its fragments C$_{58}^+$ and C$_{56}^+$. The peaks for C$_{50}^+$ and C$_{62}^+$ are present on the shoulders. Both of these spectra were obtained after ∼ 10$^4$ s of ion bombardment with $\alpha \sim 80°$ and $\theta_{DR} = 87.8°$.

Figures 5a and 5b are for heavily irradiated target with 100 keV $Kr^+-C$ and $Xe^+-C$, respectively. Both of these spectra were obtained after $\sim 10^4$ s of ion bombardment at $\alpha = 80°$ which implies sputtering away of $10^3-10^4$ surface layers. Figure 5a is for $Kr^+-C$ and is dominated by $C_{70}^+$ and its fragments $C_{68}^+$ and $C_{66}^+$ along with shoulders of $C_{82}^+$, $C_{60}^+$ and $C_{50}^+$ are present but with reduced intensities. The $Xe^+-C$ spectrum of Figure 5b is likewise dominated by $C_{60}^+$ and its fragments $C_{58}^+$ and $C_{56}^+$. The peaks for $C_{50}^+$ and $C_{62}^+$ are present on the shoulders. The main peak in both the cases appears at $C_{m-2}^+$.

## 4 Discussion

Heavy ion induced physical and chemical processes in graphite have been seen leading to cluster formation as well as their subsequent fragmentation. It can be shown that the heavy ion irradiation has three essential features which make it useful for the study of carbon clusters especially the fullerenes:

- Almost all existing bonds between carbon atoms along the ion path are broken on the time scale $\sim 10^{-14}-10^{-15}$ s [14]. For example a 100 keV $Xe^+$ ion has a range of $\sim 660$ Å and takes $\sim 10^{-13}$ s to deposit its energy before coming to a stop. The primary recoil energy distribution $\propto E_r^{-3/2}$ [15]. For $Xe^+-C_1$ the carbon recoil energy $E_r(C_1)$ varies from $\sim E_b$ at $\theta_r(max) \approx \pi/2$ to 15:4 keV for $\theta_r(min) \approx \pi/4$. This corresponds to the range of projectile's scattering angles between $\varphi_{min} \sim 0.1°$ and $\varphi_{max} \{ sin^{-1}(m_2/m_1)\} = 5.24°$. Since the majority of these ion-target atom collisions favour low energy recoils, the forward moving recoils in these cones have half angles $\theta_r > \pi/4$ in the case of $Xe^+-C_1$. The cone density along the track follows from the primary recoils' energy distribution $\propto cos_{\theta_r}^3$. Thus the lower $E_r$ and high $\theta_r$ primaries are wrapped around the ion path with a constant linear density.
- The energy of these primaries is further distributed in the $C_1-C_1$ collisions with the characteristic scattering and recoil angle $= \pi/2$ and cascading of collision events with a linear recoil density $\propto E_r^{-2}$. In heavy ion bombarded metallic targets the characteristic time scales of the cascade $\tau_{cascade} \sim 10^{-12}$ s [14]; in insulators and semiconductors it could take an order of magnitude longer time.
- Direct Recoils that originate in binary collisions between incident ions and the surface constituents carry with them the information that characterizes the dynamics of cluster formation as well as any subsequent fragmentation under continuous ionic bombardment.

For the present series of experiments the ion energy, angle of incidence with surface normal and the direct recoil angles were so chosen that heavy ions deposit considerable amounts of energy

~40 − 160 [$e^-$/Å] in a restricted volume of not more than 30 atomic layers beneath the surface by choosing $\alpha = 80°$ and with the ion penetration depth ~ 660−1300 Å at 100 keV. The incident heavy ion flux ensures that in addition to the generation of intense collision cascades, maximum randomization of the target constituents occurs.

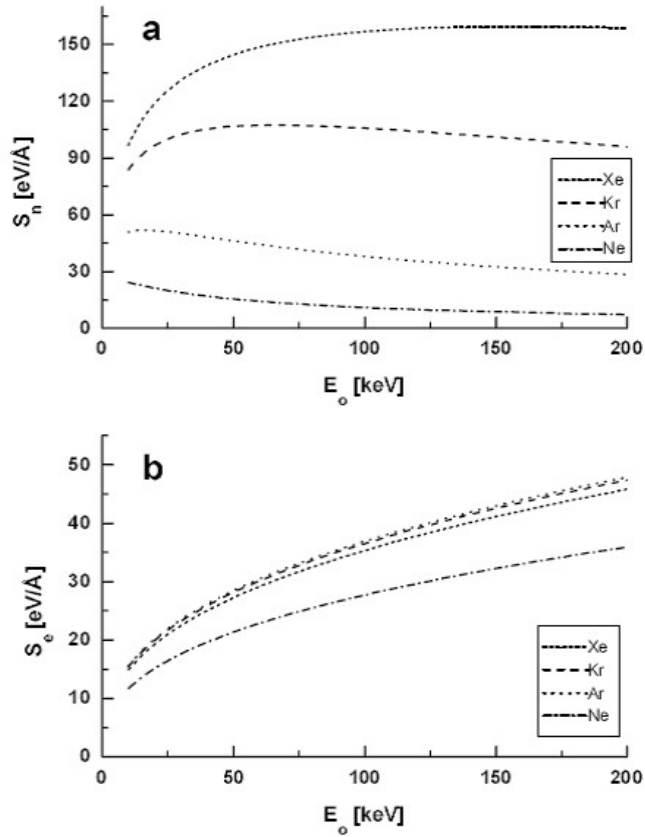

Fig. 6. Shows heavy ion stopping powers $S_n$ (nuclear) in Figure 6a and $S_e$ (electronic) in Figure 6b, respectivey, by using TRIM96 code [16]. The graphs are presented in Figure 6 for the 4 categories of ions Ne$^+$, Kr$^+$ and Xe$^+$ in the energy range of 10 to 200 keV.

Two spatial regions are therefore, generated by the incident ions on graphite in different time regimes. The first being a cylindrical region with length equal to the respective ion range and the radius ~ few to tens of _A is created by each incident ion on a time scale ~$10^{-13}$ s as discussed above. Ion's range determines the length of the cylinder, whereas radius is determined by the low energy recoils. This high density region seems to provide the necessary environment for the evolution of energetically the most favourable cluster population. The second is a slowly expanding sphere of increasing number density $\propto E_r^{-2}$ and is dependent on nuclear stopping by the constituents of the energy dissipating medium. This occurs on a slower time scales ~ $10^{-11}$ s.

The ion induced physical and chemical changes can be envisaged by considering heavy ion stopping powers $S_n$(nuclear) and $S_e$(electronic) by using TRIM96 code [16]. These are presented in Figure 6 for the 4 categories of ions $Ne^+$, $Ar^+$, $Kr^+$ and $Xe^+$ at 100 keV, respectively. It can be seen from corresponding values of $S_n$ and $S_e$ that the heavier the projectile higher is the nuclear stopping powers and therefore, the smaller shall be the ion range. $Ne^+$ incident on C1 is shown only for the sake of comparison. The ratio $S_n/S_e$ is 4.44, 2.9 and ~ 1 case of $Xe^+$, $Kr^+$ and $Ar^+$ on $C_1$. We believe that these ratios have a direct bearing on the energetics of C−C bond formations as well as fragmentation processes.

The other significant feature of $S_n$ vs. $S_e$ is the fragmentation of clusters. These clusters may have been evolved earlier due to the mechanisms suggested before. Subsequent bombardment brings in the fragmentation or partial bond-breaking sequences into play. The dominance of $C_{70}$'s fragments in $Kr^+$−$C_m$ and those of $C_{60}$'s in $Xe^+$−$C_m$ spectra at $\theta_{DR}$ =87.8° show that the direct recoiling constituents of the spectra are due to cluster fragmentation according to $\boldsymbol{C_m \to C_{m-2} + C_2}$. Similar observations have been made by others [17, 18] in collision induced fragmentation of fullerenes by energetic ions or those noted in the output of electron impact ion sources [19]. The $C_2$ loss observation has been confirmed by various groups in photo fragmentation experiments [20, 23].

The energy spectra reveal linear and chain complexes as well as fullerenes (m ≥ 20). The results clearly identify peaks corresponding to various clusters and their respective fragments which follow from the ion induced fragmentation processes. The presence of charged fullerenes in the direct recoil spectra implies their pre-existence prior to the primary knock-on collision. The intriguing aspect is the difference in the resultant dominant species C in the two heavily bombarded cases of $Kr^+$ and $Xe^+$ on C. This information may help us to understand the structure and composition of the heavily irradiated amorphous graphite surface and the mechanisms of direct recoiling of carbon clusters.

# 5 Conclusions

Carbon clusters have been generated and detected by a novel technique of Direct Recoil measurements under energetic heavy ion bombardment of amorphous graphite. The evolution of clusters and their subsequent fragmentation under continuing ion bombardment is revealed by detecting various clusters in the energy spectra of the Direct Recoils emitted as a result of collision between ions and surface constituents. Present work is an addition to the already existing techniques for the studies of clusters. It, however, introduces an element of control in the form of ion species and its energy for the monitoring of the cluster energetics.